\documentstyle[12pt,aasms4]{article}
\def\la{\mbox{\raisebox{-0.1ex}{$\scriptscriptstyle \stackrel{<}{\sim}$\,}}}
\def\ga{\mbox{\raisebox{-0.1ex}{$\scriptscriptstyle \stackrel{>}{\sim}$\,}}}

\newcommand{\Fig}{\mbox{\sc Fig. }}
\newcommand{\md}{\mbox{${ m_d }$\,}}
\newcommand{\mb}{\mbox{${ m_b }$\,}}
\newcommand{\mt}{\mbox{${ m_t }$\,}}
\newcommand{\mr}{\mbox{${ m_r }$\,}}
\newcommand{\mlin}{\mbox{${ m_{lin} }$}}

\newcommand{\rs}{\mbox{${ r_s }$\,}}
\newcommand{\rsm}{\mbox{${ r_s ^m }$}}
\newcommand{\pr}{\mbox{$ P_{r_s} $\,}}
\newcommand{\nep}{${\rm N_{ep}}$\,}
\newcommand{\tsp}{${\rm T_{sp}}$\,}
\newcommand{\nref}{\mbox{${\rm N_{ref}}$\,}}

\newcommand{\deltvobs}{\mbox{${\rm \delta t _{vobs}}$}}
\newcommand{\cn}{\mbox{${\rm C_n^2}$\,}}
\newcommand{\avcn}{\mbox{$\overline {\rm C_n^2}$}}
\newcommand{\kolind}{\mbox{$ { 11 \over 3 } $}}
\newcommand{\dmu}{$ {\rm pc ~ cm ^ {-3} } $\,}
\newcommand{\dtnu}{$ {\rm sec ~ kHz ^ {-1} } $\,}
\newcommand{\phir}{\mbox{$ {\rm \phi _{ref} } $\,}}

\newcommand{\refr}{\mbox{$ {\rm \theta _{ref} } $\,}}
\newcommand{\rmsref}{\mbox{$ {\rm \delta \theta _{ref} } $\,}}

\newcommand{\diff}{\mbox{$ {\rm \theta _{diff} }$\,}}

\newcommand{\nd}{$ \nu _d $\,}
\newcommand{\ndo}{$ \nu _{d_o} $\,}
\newcommand{\ndm}{$ \nu _d ^m $ }
\newcommand{\ndu}{\mbox{ $ \nu _{d_c} $ }}
\newcommand{\ndc}{\mbox{ $ \nu _{d_c} $ }}

\newcommand{\td}{$ \tau _d $\,}
\newcommand{\tdm}{ $ \tau _d ^m$ }
\newcommand{\Fm}{ $ {\rm F} ^m$ }

\newcommand{\dtn}{\mbox{ $ dt / d \nu $ }}

\newcommand{\viss}{\mbox{ $ V_{iss} $ }}
\newcommand{\vep}{\mbox{ $ V_{obs _{\bot }} $}}
\newcommand{\smod}{\mbox{$ \sigma _{mod} $\,}}
\newcommand{\sest}{\mbox{$ \sigma _{est} $\,}}

\newcommand{\svirr}{\mbox{$ \sigma _{v,irr} $\,}}
\newcommand{\svobs}{\mbox{$ \sigma _{v,obs} $\,}}
\newcommand{\scal}{\mbox{$ \sigma _{cal} $}}
\newcommand{\spol}{\mbox{$\sigma _{pol}$\,}}
\newcommand{\sn}{\mbox{$\sigma _{n}$\,}}
\newcommand{\xn}{\mbox{$x_n$\,}}
\newcommand{\xiss}{\mbox{$ x_{iss} $\,}}
\newcommand{\xobs}{\mbox{$ x_{obs} $\,}}
\newcommand{\xmod}{\mbox{$ x_{mod}$\,}}

\newcommand{\AandB}{$ \{ A , B \} $\,}
\newcommand{\bwtime}{$ \{ \nu _d , \tau _d \} $\,}
\newcommand{\bwflux}{$ \{ \nu _d , F \} $\,}
\newcommand{\timeflux}{$ \{ \tau _d , F \} $\,}

\newcommand{\AB}{$ r_s \{ A , B \} $\,}
\newcommand{\BT}{$ r_s \{ \nu _d , \tau _d \} $\,}
\newcommand{\BF}{$ r_s \{ \nu _d , F \} $\,}
\newcommand{\TF}{$ r_s \{ \tau _d , F \} $\,}
\newcommand{\BD}{$ r_s \{ \nu _d , d t / d \nu \} $\,}
\newcommand{\BcT}{$ r_s \{ \nu _{d_c} , \tau _d \} $\,}
\newcommand{\BcF}{$ r_s \{ \nu _{d_c} , F \} $\,}
\newcommand{\BcD}{$ r_s \{ \nu _{d_c} , d t / d \nu \} $\,}
\newcommand{\bwctime}{$ \{ \nu _{d_c} , \tau _d \} $\,}
\newcommand{\bwcflux}{$ \{ \nu _{d_c} , F \} $\,}
\newcommand{\bwcdrift}{$ \{ \nu _{d_c} , d t / d \nu \} $\,}
\newcommand{\bwtimemod}{$ \{ \nu _d ^m , \tau _d ^m \} $\,}
\newcommand{\bwfluxmod}{$ \{ \nu _d ^m , F ^m \} $\,}
\newcommand{\timefluxmod}{$ \{ \tau _d ^m , F ^m \} $\,}
\newcommand{\norma}{\mbox{$ {\rm PSR ~ B0823+26(I) } $ }}
\newcommand{\normb}{\mbox{$ {\rm PSR ~ B0823+26(II) } $ }}
\newcommand{\egta}{\mbox{$ {\rm PSR ~ B0834+06(I) } $ }}
\newcommand{\egtb}{\mbox{$ {\rm PSR ~ B0834+06(II) } $ }}
\newcommand{\egtc}{\mbox{$ {\rm PSR ~ B0834+06(III) } $ }}
\newcommand{\egtd}{\mbox{$ {\rm PSR ~ B0834+06(IV) } $ }}

\newcommand{\elevb}{\mbox{$ {\rm PSR ~ B1133+16(II) } $ }}

\newcommand{\ninea}{\mbox{$ {\rm PSR ~ B1919+21(I) } $ }}

\begin{document}


\title{\LARGE\bf Long-Term Scintillation Studies of Pulsars: \\
III. Testing Theoretical Models of Refractive Scintillation}

\vspace{5.0cm}
 
\author{\bf N. D. Ramesh Bhat\footnote{send preprint requests to $ bhatnd@ncra.tifr.res.in $}, 
A. Pramesh Rao, and Yashwant Gupta }


\begin{center}
{National Centre for Radio Astrophysics, Tata Institute of Fundamental Research, \\
Post Bag 3, Ganeshkhind, Pune - 411 007, India}
\end{center}
 
\vspace{5.0cm}
 
\begin{center} {\bf Accepted for publication in The Astrophysical Journal} \end{center}
 

\begin{abstract}

Refractive interstellar scintillation (RISS) is thought to be the cause behind a variety of phenomena 
seen at radio wavelengths in pulsars and compact radio sources.
Though there is substantial observational data to support several observable consequences of it, the
quantitative predictions from theories have not been thoroughly tested.
In this paper, data from our long-term scintillation study of 18 pulsars in the DM range 3$-$35 \dmu are used
to test the relevant theoretical predictions.
The variabilities of decorrelation bandwidth (\nd), scintillation time scale (\td) and flux density (F)
are examined for their cross-correlation properties and compared with the existing predictions.
The theory predicts a strong correlation between the fluctuations of \nd and \td, and strong
anti-correlations between those of \nd and F, and \td and F.
For 5 pulsars, we see a reasonable agreement with the predictions.
There is considerable difficulty in reconciling the results for the rest of the pulsars,
most of which show the positive correlation between \nd and \td, but are characterized by poor flux
correlations.
In general, the measured correlations are lower than the predicted values.
Our analysis shows that while the underlying noise sources can sometimes reduce the degree of correlation, 
they cannot give rise to an absence of correlation.
It is also unlikely that the observed poor flux correlations arise from a hitherto unrecognized form of 
intrinsic flux variations of pulsars.
For PSR B0834+06, which shows anomalous behaviour in the form of persistent drift slopes, positive 
correlation is found between \td and the drift-corrected \nd.
Many pulsars show an anti-correlation between the fluctuations of \nd and the drift rate of intensity
patterns, and this is in accordance with simple minded expectations from theory.
The detections of correlations between the fluctuations of different observables, and a reasonable agreement 
seen between the predicted and measured correlations for some pulsars confirm RISS as the primary cause of 
the observed fluctuations. However, the complexity seen with the detailed results suggests the necessity 
of more comprehensive theoretical treatments for describing refractive fluctuations and their 
cross-correlations.

\end{abstract}

{\it Subject headings: }
{ISM:General -- Pulsars:General -- Radio continuum:ISM -- Scattering}


\section{Introduction }

With the recognition of interstellar propagation effects in the long-term pulsar flux variations
(Sieber 1982), a new class of scintillation, namely Refractive Interstellar Scintillation (RISS), 
emerged in radio astronomy (Rickett, Coles \& Bourgois 1984).
Since then, much progress has been made, both in theoretical as well as observational fronts, 
to understand this new form of scintillation.
RISS is thought to arise from propagation through large-scale (scale sizes $\ga$ $10^{11}$ m) electron 
density inhomogeneities in the interstellar medium (ISM) and hence forms a powerful tool to probe the 
ISM at such scales (see Rickett (1990) for a review).
The growing interest in RISS over the recent years is largely motivated from its applicability beyond the
field of pulsars and valuable insights it provides on the distribution of electron density irregularities 
in the ISM.
RISS is also investigated with a view to distinguish between the intrinsic and extrinsic 
effects on signals from pulsars and compact radio sources, and for an improved understanding of strong 
scattering phenomena at radio wavelengths.
Most investigations of RISS effects have been, so far, largely based on observations of pulsars.
Due to their high spatial coherence and pulsed nature of radiation, pulsars are expected to 
exhibit a variety of observable effects due to RISS.

Various observable effects of interstellar scattering (ISS) on pulsar signals can be classified into
(i) diffractive effects, (ii) refractive effects and (iii) combined effects due to diffraction and 
refraction.
Detailed theoretical treatments can be found in Cordes, Pidwerbetsky \& Lovelace (1986)
and Romani, Narayan \& Blandford (1986) 
(also see Rickett (1990), Narayan (1992) and Hewish (1992) for reviews).
The refractive effects include, in addition to the familiar long-term 
(days to weeks at metre wavelengths) flux variations, modulations of Diffractive Interstellar 
Scintillation (DISS) observables, drifting of intensity patterns, timing perturbations and image wandering.
The combined effects of diffraction and refraction in the ISM can, occasionally, give rise to dramatic
events, in the form of periodic intensity modulations in time and frequency, which form potential tools 
to probe pulsar magnetospheres (e.g. Wolszczan \& Cordes 1987).
RISS is also thought to be the cause behind slow (months to years) flux variations (mainly at metre 
wavelengths) seen with a number of compact extra-galactic radio (EGR) sources, and unusual flux variations, 
known as extreme scattering events (ESE), seen with some quasars (at centi-metre wavelengths)
(e.g. Bondi et al. 1994; Spangler  et al. 1993; Fiedler et al. 1987, 1994).
However, unlike DISS, there are several aspects of RISS phenomena which remain to be well
understood (cf. Rickett 1990; Cordes, Rickett \& Backer 1988).

Properties of various kinds of observables of interest in the ISS of pulsars are discussed by
Cordes et al. (1986) and Romani et al. (1986).
The effects investigated by them include modulations of diffraction patterns, fluctuations in 
pulse arrival times, drifting bands in dynamic spectra, multiple imaging, angular wandering 
and distortion of scattered images, and the well known long-term flux variations.
In the formalism of Romani et al. (1986), refraction effects are treated as weak perturbations of a bundle 
of rays scatter-broadened by small-scale density inhomogeneities.
The small- and large-scale inhomogeneities are assumed to be part of a power-law form of spectrum, and the
scattering model considered is a thin screen placed between the source and the observer.
They extend this formalism to compute the auto- and cross-correlation functions of different observables
analytically. The explicit predictions of cross-correlations given by them for the fluctuations of flux and 
DISS observables (\nd and \td) are particularly suitable for experimental verifications.
Cordes et al. (1986) consider a phase screen comprising the `slow' and `rapid' components of phase
fluctuations, and employ a formalism based on the Kirchoff diffraction integral. 
Refractive effects are analyzed using a Taylor expansion of the `slow' component.
They do not make any quantitative predictions of correlation properties, but the formalism can be 
extended to make the relevant predictions.

Although models based on simple power-law spectra for density fluctuations and a single phase screen 
appear to be simplistic to describe much more complicated scenarios of ISS, it is worthwhile testing the
explicit predictions of such models due to the following reasons. Firstly, cross-correlations between the 
fluctuations of observables form more direct and rigorous tests of the RISS theory. Secondly, it is 
necessary to critically examine to what extent the observations support the predictions of simple models, 
the results from which would provide valuable inputs for suitable refinements of the models.

A number of observational attempts have been made in the recent past towards investigating basic
consequences of RISS such as long-term flux variations of pulsars (e.g. Kaspi \& Stinebring 1992; 
Gupta, Rickett \& Coles 1993) and changing form of pulsar dynamic spectra (e.g. Gupta, Rickett \& 
Lyne 1994; Bhat, Rao \& Gupta 1998a).
But only few observational studies have been made to verify the predicted cross-correlation properties 
between the scintillation observables. 
To the best of our knowledge, there have been only two instances of such attempts. 
Recent timing observations of the milli-second pulsar PSR B1937+21 
yield some evidence supporting the theoretical predictions
(Lestrade, Cognard \& Biraud 1995). 
An anti-correlation between the flux variations and times-of-arrival is reported, which is found to be
consistent with the prediction for a density spectrum with power-law index $ \alpha  < 4 $.
Also, observations of PSR B0329+54 by Stinebring, Faison \& McKinnon (1996) showed that the correlation 
properties between variations of flux, decorrelation bandwidth and scintillation time are in agreement 
with the theoretical predictions. 
Given the complexity of ISS, these examples may not necessarily represent the typical behaviour.
It is highly desirable to test the predictions for a large number of pulsars so that the possibility 
of an observational bias can be reduced. Such extensive tests have not been carried out so far.

We have carried out a long-term, systematic study of the scintillation properties of eighteen pulsars 
using the Ooty Radio Telescope (ORT) at 327 MHz during a three-year period from 1993 to 1995. 
One of the prime objectives was to study refractive effects in pulsar scintillation.
The dynamic scintillation spectra of pulsars were regularly monitored for 10 to 90 epochs 
over time spans $ \sim $ 100 to 900 days, from which scintillation parameters, viz., decorrelation
bandwidth (\nd), scintillation time scale (\td), frequency drift rate (\dtn) and pulsar flux density 
(F) were measured.
Details of observations and results on diffractive and refractive scintillation properties have been 
discussed in an earlier paper (Bhat et al. 1998a, hereinafter referred to as Paper I). 
The results were also used to study the distribution of scattering material in the local interstellar 
medium (LISM) (Bhat, Gupta \& Rao 1998) and to investigate the form of the electron density spectrum 
in the ISM (Bhat, Gupta \& Rao 1998b, hereinafter referred to as Paper II). 
The results from a cross-correlation analysis of the fluctuations of various parameters are presented 
in this paper.
We give a brief review of the relevant theory in \S 2.
Our data analysis and results are presented in \S 3.
In \S 4, we compare the results with the theoretical predictions.
In \S 5, we discuss some of the possibilities of reconciling the observations and theories.
Our conclusions are summarized in \S 6.


\section{Theoretical Background}

Although several papers have addressed different issues related to refractive scintillation effects,
we find that two papers, Cordes et al. (1986) and Romani et al. (1986), contain theoretical treatments 
relevant to our investigations.
Despite the differences in the details, both papers attempt to explain a variety
of effects on scintillation observables that can be caused by RISS.
The authors rely on ray tracing techniques to explain refractive effects and wave techniques for 
diffractive effects.
The scattering model considered is that of a thin screen placed between the source and the observer.

Cordes et al. (1986) analyzed refractive effects using a Taylor expansion of the ``slow'' component of
phase perturbations, sometimes termed as the `refractive phase' \phir, which varies over length scales 
$\sim$ ``multi-path scale'' ($ r_{mp} \sim Z \ \diff $, where $Z$ is the distance to the 
scattering screen and \diff is the diffractive scattering angle).
The refraction-induced fluctuations of scintillation parameters are treated in terms of `gradient' and 
`curvature' components of this refractive phase. The observables such as spatial scale ($ s_d $) and
characteristic bandwidth ($ \nu _d $) of scintillation patterns, the mean flux density (F) and the drift 
rate (\dtn) of patterns are expressed in terms of refraction angle, \refr, and a ``gain'' term, G, which 
are defined as 

\begin{equation}
\refr = { \lambda \over 2 \pi } ~ \left( { \partial \phir \over \partial r } \right)
\hspace{1.0cm} 
{\rm and }
\hspace{1.0cm} 
G = \left( 1 + { \lambda \ Z \over 2 \pi } {\partial ^2 \phir \over \partial r^2 } \right) ^{-1}
\end{equation}



According to the relations given by Cordes et al. (1986), the refractive modulations of spatial scale 
(or, alternatively, the scintillation time scale, \td) and flux density (F) are governed by the gain 
term ($G$) 
whereas the decorrelation bandwidth (\nd) is modulated by both \refr as well as $G$. 
Modulations of the drift slope are due to variations of \refr.
Therefore, it is reasonable to expect variabilities of diffractive scintillation parameters and the flux
density to be inter-related to each other. 
For example, larger values of $G$ increase the apparent flux, whereas the scintillation time is reduced 
by a similar factor.
The variability of \nd is expected to be rather complicated, as demonstrated by their numerical 
simulations.
However, the authors do not make any quantitative, verifiable predictions of correlation properties of
these parameters.

Romani et al. (1986) treat refraction-induced fluctuations as
weak perturbations of a bundle of rays scatter broadened by the short-wavelength electron density 
inhomogeneities.
The source image is assumed to arise from a Gaussian-shaped bundle of such rays, which get focused, defocused
or steered by the density profiles over length scales $ \sim $ Z\diff 
(which is also the `image size' on the phase screen).
As in the work of Cordes et al. (1986), large-scale phase variations give rise to a refractive 
bending angle, \refr. 
The assumption of weak perturbations essentially means the refractive displacements of the ray bundle,
given by X $ \sim $ Z\refr, are small compared to the image size, Z\diff; which, in turn, implies 
a refractive bending angle that is smaller than the small-scale scattering angle (\refr $ < $ \diff).
Based on this formalism, fractional fluctuations of observables of interest, such as the source size
$ \Omega $, \nd, \td, F, etc., are calculated by integrating ``intensity-weighted'' \phir over the 
image size (see Blandford \& Narayan (1985) for a detailed discussion).
The weighting function depends on the observable.
The mean auto- and cross-correlations are computed for the observables using the Fourier transform 
method (see Appendix A of Romani et al. (1986) for details).
The authors group the variabilities of scintillation parameters into two classes, viz., 
(i) curvature-induced and (ii) gradient-induced, and argue that, in general, 
one can expect the quantities belonging 
to a particular class to co-vary among themselves and the quantities of different classes to show 
dissimilar variations.
They emphasize that fluctuations of three quantities, viz., decorrelation bandwidth, scintillation time 
and flux are most suitable for observational verification and suggest that detections of such fluctuations,
particularly their cross-correlations, will help towards understanding the refractive scintillation effects in the ISM.

In this paper, we restrict ourselves to the observables that can be obtained from our data.
Out of 5 quantities which can be measured from the dynamic spectrum at a given epoch
$-$ \nd, \td, \dtn, F and the scintillation index (\md), there are clear predictions made by 
Romani et al. (1986) for the cross-correlation properties of \nd, \td and F.
It should be mentioned that the scattering model considered by them employs a single phase screen and 
a simple power-law description for the underlying density fluctuations.
This may be an idealized simplification of the real situation.
We have summarized these predictions in Table 1 in terms of zero-lag correlation coefficients
for 3 different types of density spectra with power-law indices ($ \alpha $) of \kolind, 4 and 4.3.
In general, the observable effects of refractive scintillation are thought to be strongly dependent 
on the form of the density spectrum, but the predicted correlation properties between the fluctuations of 
\nd, \td and F change only marginally with $ \alpha $. 
To summarize their predictions, a high positive correlation (0.75 to 0.79) is expected between variations of 
decorrelation bandwidth and scintillation time whereas high anti-correlations  are 
expected between variations of decorrelation bandwidth and flux ($-0.76$ to $-0.84$) and those of 
scintillation time and flux ($-0.50$ to $-0.64$).


\section{Data Analysis and Results}

The theoretical predictions described above can be tested using the pulsar scintillation data presented 
in Paper I.
Time series of four parameters $-$ decorrelation bandwidth (\nd), scintillation time scale (\td), 
drift rate of intensity patterns (\dtn) and flux density (F) $-$ for each pulsar are presented in 
Figs 4(a)$-$(x) of Paper I.
Our observations show large-amplitude fluctuations of these quantities; 
typically $40-50$\% fluctuations for \nd and F, 
$20-30$\% for \td, 
and rms fluctuations of a few \dtnu for \dtn (Paper II).
The dynamic spectra are found to vary significantly over time scales as short as $2-3$ days, 
which, to first order, is in accord with basic expectations of refractive modulations at our 
observing frequency (Paper I).
However, while the modulations of scintillation observables \nd, \td, \dtn and pulsar flux density
(F ) due to refractive scintillation effects in the ISM have been detected, 
the depths of modulations are found to be much larger than the expectations of a 
Kolmogorov-type density spectrum (see Paper II).
In this paper, we present a correlation analysis of this data set and examine how well the results 
are in agreement with the existing theoretical predictions.

In addition to studying the correlation properties between the fluctuations of \nd, \td and F, 
for which theoretical predictions exist, we also examine the correlation properties between the 
fluctuations of \nd and \dtn.
The motivation for this is as follows.
According to the current models of refractive scintillation, the `apparent' or `instantaneous' 
decorrelation bandwidth (\nd) is expected to be less than the characteristic bandwidth (\ndo) 
(ie., \nd in the absence of refraction effects) when the refractive bending angle is significant
(cf. Cordes et al. 1986; Gupta et al. 1994). 
Our observations (Paper II) show that there are considerable variations of \refr (rms refractive angle 
\rmsref $ \sim $ 1 mas)  over the observing time spans (typically $ \sim $ 100 days).  
Therefore, it is reasonable to expect variations of \refr (or, alternatively, the gradient component of 
the slowly varying phase \phir) to play a substantial role in the long-term modulation characteristics 
of \nd.
To examine the possible connection between the two effects, we carry out a correlation analysis between 
\nd and \dtn.
Since the modulations of \nd are insensitive to the sign of the drift slope (see Gupta et al. (1994) 
and Cordes et al. (1986) for details), we correlate variations of \nd with those of $ | \dtn | $.

The pulsars and the periods of observation are tabulated in columns (2) and (3) of Table 2.
Columns (4) and (5) of Table 2 give the number of epochs of observation (\nep) and the time 
span of data (\tsp) respectively.
For pulsars PSR B0823+26, PSR B0834+06, PSR B1133+16 and PSR B1919+21, there are multiple entries,
which correspond to data from different observing sessions.
The symbols I$-$IV, when attached alongside pulsar names, indicate the data from a given session
(see Tables 1 and 2 of Paper I for details).
Though data from each session for these pulsars span a large number of refractive time scales (as per our 
expected estimates), the results show that the diffractive and refractive scintillation 
properties sometimes vary significantly between successive sessions (see Papers I and II).
Hence we treat them as independent data sets for the correlation analysis.

Details concerning the statistical quality of our data and possible non-ISS effects which give rise to 
the modulations of the measured parameters are described in Appendices A and B of Paper II.
The correlation properties are meaningful only when the data span several refractive time scales.
Rough estimates of the number of refractive cycles of fluctuations (\nref) are given in Table 3 of Paper II.
We find that for pulsars PSR B2016+28, PSR B1540$-$06 and PSR B2310+42, the data span only a few 
time scales of fluctuations ($ \nref \sim 3 $) and hence their correlation properties 
may not be reliable.
Also, along the directions of PSRs B1540$-$06 and B1604$-$00, 
variation in the transverse component of the Earth's orbital motion (\vep ) 
contribute substantially to the modulations of scintillation time scale (\td).
The expected fractional variation in \td due to this (\deltvobs) 
is comparable to the measured modulation index of \td ($ie., \deltvobs \approx \mt$).
Therefore, the \td correlations (i.e., between \nd and \td, and \td and F) of these pulsars may not be
meaningful.
Due to their comparatively large fractional linear polarizations ($\mlin$ $\sim$ 0.6 to 0.8 at 400 MHz),
the flux density modulations (as measured by ORT) of PSRs B1237+25 and B1929+10 can get significantly 
modified by the variable Faraday rotation (due to the Earth's ionosphere).
Therefore, the flux correlations of these pulsars, 
(i.e., between \nd and F, and \td and F) may not be reliable.
Although, we carry out the correlation analysis for all the 18 pulsars, the results of above-mentioned
pulsars should be treated with caution.
For the rest of the pulsars, these three effects are not significant. 


\subsection{Cross-correlation Analysis and Results}

Sample data from our observations are shown in Figs. 1(a)$-$(l) and 2(a)$-$(c) in the form of scatter plots
of various combinations: \nd vs \td, \nd vs F, \td vs F and \nd vs \dtn.
The trends visible therein give some idea of the nature of correlations between different combinations.
A full cross-correlation function analysis is not practical with our data due to the limited number of 
measurements in the time series of the parameters.
Therefore, we compute cross-correlation coefficients, i.e., normalized zero-lag values of cross-correlation
functions, for which predictions are readily available.
The quantity we use as a measure of the correlation between the parameters is 
the {\it Spearman rank-order correlation coefficient} ($ r_s $), which is defined as 
(e.g. Press et al. 1992)

\begin{equation}
r_s ~ = ~ { \sum _i (R_i - \overline R) ~ (S_i - \overline S) \over 
\sqrt { \sum _i (R_i - \overline R)^2 } ~ \sqrt { \sum _i (S_i - \overline S)^2 } }
\end{equation}

\noindent
where $ R_i $ and $ S_i $ represent the ranks of the two quantities $ x_i $ and $ y_i $ for which the
correlation coefficient is computed and the summation is carried out over the total number of available
data points in the time series. 
The quantities $\overline R$ and $\overline S$ denote the average values of $ R_i $ and $ S_i $
respectively, over the entire time span of observation.
The rank correlation method is preferred over the normally used linear correlation due to the following 
reasons. 
It is a non-parametric test, where no assumptions are made about the distributions of the quantities.
A non-parametric correlation is more robust than the linear correlation method.
In addition, the rank correlation method is less sensitive to outliers than the linear 
correlation, where they are likely to introduce a bias in the mean, thereby giving rise to unreliable 
correlations.
The rank correlation test is also applicable in the case of non-linear dependence between the quantities,
unlike the case for linear correlation.

We also computed the linear correlation coefficients (also referred to as ``Pearson's r'';
Press et al. 1992), which sometimes differ substantially from their rank correlation counterparts, 
and in general have poorer confidence intervals.
The results from the two methods, however, are found to be in qualitative agreement.
Other methods of examining the relation between two quantities, such as {\it Kendall's tau} 
(Press et al. 1992), also give qualitatively similar results.
Therefore, we restrict ourselves to the results obtained from the rank correlation technique 
in the subsequent discussion.

The confidence intervals of the rank correlation coefficients are derived using the ``bootstrap'' method
(Efron 1979; Diaconis \& Efron 1983).
The procedure is as follows.
From the observed sample of \nep data points in a given data, a new sample of \nep values is generated by a
random re-arrangement process.
The rank correlation coefficient is computed for this bootstrap sample.
This process is repeated a very large number of times and a probability distribution is created from the
coefficients obtained for all the samples.
We have made use of $ 10^5 $ bootstrap samples generated from the basic data set.
Computations for much larger number of samples produce very little changes in the statistical 
properties of the distribution.
The desired confidence intervals are obtained by appropriately integrating over this distribution curve.
For instance, if \pr is the desired confidence level in percentage, then the distribution is integrated from either 
of the wings inward to $(100-\pr)/2$ percent of the total area under the distribution curve to get the 
corresponding limits of the confidence interval.
The analysis is carried out for a wide range of intervals (\pr \ ranging from 60\% to 95\%).
Due to the limited number of measurements (\nep) and finite number of independent refractive cycles 
(\nref) spanned by our data, we prefer 90\% levels to decide the significance of our correlation results.
The estimated width of the confidence interval reflects, to first order, the statistical quality of our 
data in terms of \nep and \nref, and the uncertainties in individual measurements.
It is possible that this method still underestimates the uncertainty when the data do not span many
independent cycles of fluctuations and/or the amplitudes do not span the total range of values.

The analysis described above is carried out for the following combinations of the observables:
(i) decorrelation bandwidth and scintillation time scale (\nd vs \td),
(ii) decorrelation bandwidth and flux density (\nd vs F),
(iii) scintillation time scale and flux density (\td vs F), and
(iv) decorrelation bandwidth and drift rate (\nd vs \dtn).
The results are tabulated in columns (3), (4), (5) and (6) respectively of Table 3.
Pulsars for which our correlation results may not be reliable, due to reasons described earlier,
are indicated by an asterisk symbol.
The confidence intervals of the correlation coefficients obtained from the bootstrap method are shown
as the subscript (90\% lower limit) and the superscript (90\% upper limit).
Throughout this paper, we use \AandB to refer to the cross-correlation between the quantities A and B,
and \AB to denote its rank correlation coefficient.

Our analysis shows that the correlation properties change significantly from pulsar to pulsar, 
and the coefficients vary over a very wide range.
Excluding the unreliable correlations (see Table 3),
\BT ranges from $-0.14$ (\egtb) to 0.83 (\elevb), 
\BF from $-0.57$ (PSR B1508+55) to 0.45 (PSR B0329+54) and 
\TF from $-0.86$ (PSR B1508+55) to 0.37 (\egta).
The values of \BD range from $-0.60$ (PSR B$1604-00$) to 0.11 (PSR B1747$-$46).
There are a large number of data (roughly one third) for which the magnitudes of correlation coefficients
are very low ($ | \rs | $ \la 0.1), and these will be treated as ``insignificant correlations'' 
in our further discussion.
Significance of the rest of the correlations are decided purely on the basis of the limits of the 
confidence intervals.
In the discussion that follows, we usually refer to correlation coefficients unless a special mention is
made of the confidence limits.
In column (7) of Table 3, we indicate the nature of agreement between the correlations and the theoretical
predictions.


\section{Comparison with Theoretical Predictions}

\subsection{Correlations between DISS Observables and Flux Density}

On comparing Tables 1 and 3, we recognize that there is no pulsar for which results are in complete 
agreement with the theoretical predictions. Five pulsars show correlation properties in qualitative 
agreement with the predictions; these are PSRs B0628$-$28, B0823+26(II), B0919+06, B1508+55 and B2020+28.
However, the observed correlations are in general lower compared to their predicted values. 
For PSR B0628$-$28, results agree with the predictions within the 90\% confidence intervals, whereas 
a similar agreement is not seen with the other four for all the 3 combinations.

Correlation properties of the remaining data are rather complex. Four data sets show two of the 
combinations in qualitative agreement, but an absence or an opposite correlation for the third 
combination (these are indicated by `2A' in column (7) of Table 3).
For 11 data sets (indicated by `1A' in column (7) of Table 3), only one of the three combinations 
agrees with the prediction, while an absence (e.g. PSRs B0329+54 and B0823+26(I)) or an opposite 
correlation (e.g. PSR B0834+06(I) and B1919+21(II)) is seen for the remaining 2 combinations.
There are also examples (PSRs B0834+06(III) and B2045-16) where the fluctuations of all the three
quantities are uncorrelated to each other. Thus the results vary over a very wide range from a 
complete absence of correlations to a reasonable agreement within the confidence intervals; in between 
there are several cases where the agreement is only partial.

For PSRs B1540$-$06, B1929+10 and B2327$-$20, the observed correlations are in opposite sense to that
predicted by theory.
However, a close inspection of the relevant data reveals that these are spurious correlations arising
due to the non-randomness of the data (Figs. 4.n, 4.s and 4.x of Paper I).
PSR B1540$-$06 shows a significant fading between the first and second halves of the data;
while the measurements of the first half are consistently biased above the mean value, those of the second
half are biased below the mean.
Therefore, the observed positive correlation for \bwflux for this pulsar is not meaningful.
As mentioned in \S 3, the flux density modulations due to variable Faraday rotation is expected to be
substantial for PSR B1929+10 owing to its very large fractional linear polarization (\mlin $\sim$ 0.8 at 400
MHz; Gould 1994), which makes the observed flux correlations (\BF and \TF) unreliable.
There are some signatures of systematic trends in the time series of \nd and F of PSR B2327$-$20, which are
the probable cause behind the spurious flux correlations seen for this pulsar.
The correlations of these 3 pulsars, therefore, should not be taken seriously.

The diversity seen in the correlation properties is exemplified by Figs. 1(a)$-$(l).
PSR B0823+26(II) is an example of `reasonable agreement', 
with \BT $\approx$ 0.6, \BF $\approx$ $-0.3$ and \TF $\approx$ $-0.3$, 
and the corresponding trends are visible in Figs. 1(a)$-$(c).
For PSR B0834+06(IV) (Figs. 1(d)$-$(f)), only two combinations $-$ \BT = 0.43 and 
\BF = $-$0.23 $-$ are in qualitative agreement with the predictions.
PSR B1919+21(II) (Figs. 1(g)$-$(i)) is an example where only one combination 
$-$ \BT = 0.44 $-$ agrees with the prediction.
The extreme case of variabilities of all the 3 quantities uncorrelated to each other is shown in 
Figs. 1(j)$-$(l) through the example of PSR B2045$-$16.

\subsection{The Anti-correlation between Decorrelation Bandwidth and Drift Slope}

A fairly large number of pulsars in our sample (20 out of the 25 entries in Table 3) show an anti-correlation 
between the fluctuations of \nd and \dtn.
Such a relation can be expected from the effect of refractive bending angle on 
the intensity decorrelation in frequency (cf. Cordes et al. 1986; Gupta et al. 1994).
Figs. 2(a)$-$(c) show sample plots illustrating the anti-correlation between \nd and \dtn.
Like other combinations, correlations between \nd and \dtn also vary over a wide range:
from values as low as $ \approx $ $-$0.2 (eg: PSRs B0329+54, B0834+06(I)) to $ \approx $ $-$0.7 
(e.g. PSRs B0834+06(II) and B1133+16(III)).
If such correlations result from the `phase-gradient-induced' modulations of \nd, 
then the other bandwidth correlations such as \bwtime and \bwflux, which are expected to arise 
purely from curvature effects, may be reduced.
However, from Table 3, we see that the reduction in \BT and \BF (with respect to predicted values),
to first order, does not depend on the value of \BD.
No quantitative theoretical predictions are available for the correlation between \nd and \dtn. 
But our data suggest that there is some connection between the fluctuations of these two quantities.

\subsection{Stability of Correlations and Effect of Statistical Quality}

For pulsars PSRs B0823+26, B0834+06, B1133+16 and B1919+21, we have carried out correlation analysis 
of each distinct observing session separately, and find that the correlation properties are not stable 
between successive observing sessions. \norma shows only the \bwtime correlation, whereas all the 3 
combinations are correlated for \normb. Data of PSR B0834+06 are even more remarkable. \egta shows 
correlations between all the 3 combinations, whereas an anti-correlation between \nd and F, and 
insignificant correlations of the other 2 combinations are seen for \egtb. For \egtc, the fluctuations 
of all the 3 quantities are uncorrelated to each other, and \egtd shows an entirely different behaviour 
(\bwtime and \timeflux are positive, \bwflux is negative). For PSR B1133+16, qualitatively similar 
correlations are seen for data from sessions II and III: positive correlation between \nd and \td, and 
lack of flux correlations. However, the strength of correlation between \nd and \td (\BT) is significantly 
reduced in session III compared to that in session II (0.83 $\longrightarrow$ 0.32), which is probably due 
to an underestimation of some of the \nd values in session III (see Paper I for details). For PSR B1919+21, 
flux correlations are absent in session I, and significant in session II. 
Further, \BT is considerably larger in session II (0.21 $\longrightarrow$ 0.44).
All the 4 pulsars, however, consistently show an anti-correlation between the fluctuations of \nd and \dtn, 
albeit with varying magnitude.

In order to examine the effect of statistical quality of the data on the stability of correlation properties,
we carried out the correlation analysis for the combined data from all the observing sessions of each pulsar.
The results are given in Table 4, and the relevant scatter plots are shown in Figs. 3(a)$-$(p).
This procedure should have resulted in improved correlations and better agreement with the theoretical 
predictions, if the statistical quality of the data was the main cause for change in the correlation 
properties. On the contrary, it leads to reduced correlations, and sometimes even a reversal of the sense of
correlation. For PSR B0823+26, the flux correlations reverse the sign on combining data from the 2 sessions
(Figs. 3.b and 3.c). On combining the data from all the 4 sessions, PSR B0834+06 shows a degradation of the 
\bwtime correlation (Fig. 3.e), while retaining the positive correlation between \td and F (seen for 
sessions I and IV) (Fig. 3.g). In addition, \bwflux correlation totally vanishes (\BT = 0) (Fig. 3.f).
For PSR B1133+16, flux correlations do not turn up even with the combined data (from sessions II and III)
(Figs. 3.j and 3.k). Quite peculiar changes are seen with PSR B1919+21, where the \bwtime and \bwflux 
correlations reverse the sign (Figs. 3.m and 3.n), while the positive correlation between \td and F turns 
more significant (Fig. 3.o). Thus, improvement in the statistical quality of the data, in terms of number 
of epochs of observation (\nep) or number of refractive cycles spanned (\nref), does not result in a better 
agreement with predictions, and, therefore, is not likely to be the cause of the lack of stability seen with 
the correlation properties.


\section{ Discussion }

Although several consequences due to RISS are amply supported by our data (Papers I and II),
the cross-correlation properties between the fluctuations of various scintillation observables 
(\nd, \td, \dtn and F), do {\it not fully} agree with the existing predictions. 
Interestingly, there are several characteristics that are common for many data sets.
For example, the predicted positive correlation between decorrelation bandwidth (\nd) and scintillation
time scale (\td) are seen with 17 of the 19 statistically reliably data sets (see Table 3 for details),
whereas the general disagreement with the predictions is mainly due to poor flux correlations.
However, in contrast to the predictions, the measured correlation coefficient \BT varies over a very wide
range: from 0.21 (for \ninea) to 0.83 (for \elevb).
Further, a fairly good number of correlations (20 out of 25) are seen between the fluctuations of
decorrelation bandwidth (\nd) and the drift slope (\dtn).
A broad conclusion that can be drawn from these results is that the modulations of
the quantities \nd, \td and \dtn are somewhat in accordance with the theoretical
expectations, whilst the flux variations appear to be rather more complex.

The complexity of the results seems to be beyond the scope of simple models of refractive 
scintillation.
A satisfactory model needs to explain 
(i) the observed diverse nature of correlation properties,
(ii) poor flux correlations, and 
(iii) much wider ranges in the strengths of correlations than the existing predictions.

Our attempts at finding a possible connection between the observed correlation properties and 
properties related to scattering (such as the strength of scattering (\avcn, $u$), modulation 
indices (of \nd, \td and F), and rms refractive angle (\rmsref)) or pulsars (such as DM, distance 
and direction ($l,b$)) have not been, so far, successful.
For the given number of pulsars and the diversity seen with the correlation properties, 
individual treatments of pulsars appear rather formidable.
In this section, we examine possible ways of reconciling the results from our correlation analysis and 
the predictions from the theory.
First, we briefly address the issue of effects due to various identifiable noise sources on 
our correlation results (\S 5.1), and then examine the implications of the present data on 
intrinsic flux variations (\S 5.2).
We also consider the role of persistent drifting bands on modifying the correlation properties (\S 5.3).
Finally, on the basis of observational evidence(s) available, we emphasize the need for reconsidering some 
of the basic assumptions made by the theoretical models (\S 5.4).


\subsection{Effects due to Noise Sources}

The error bars on the sample data displayed in Figs. 1(a)$-$(l) and 2(a)$-$(c)
give some idea of the typical uncertainties due to noise sources associated 
with our measurement procedure.
These noise effects can potentially modify the correlation properties between the various quantities 
and lead to a reduction in the strength of correlation.
To understand this effect, we model the observed time series, \xobs, as a combination of 
the true variable, \xiss, and a noise source, \xn.

\begin{equation}
x_{obs} (i) ~ = ~ x_{iss} (i) ~ + ~ x_n (i)
\end{equation}

\noindent
where $x$ represents the quantity \nd, \td or F at $i^{th}$ epoch of observation.
The various kinds of noise sources relevant for each quantity
can be combined to obtain a net uncertainty (\sn), given by

\begin{equation}
\sigma _n ~ = ~ \left( \sum _{j=1} ^{j=k} \left\{ \sigma _{n,j} \right\} ^2 \right) ^{0.5} .
\end{equation}

\noindent
Here $ \sigma _{n,j}$ is the uncertainty caused by $j^{\rm th}$ noise source and $k$ is the number of
independent noise sources relevant for the quantity under consideration.
The sources of noise that are identifiable for our data include
(1) errors due to the Gaussian model fitting to the auto-correlation functions of dynamic spectra,
\smod (for \nd, \td and \dtn),
(2) statistical errors due to finite number of scintles, \sest (for \nd, \td, \dtn and F),
(3) calibration error, \scal (for F),
(4) errors due to variable ionospheric Faraday rotation, \spol (for F),
(5) errors due to the Earth's orbital motion, \svobs (for \td), and
(6) errors due to the bulk flow of the density irregularities, \svirr (for \td).
Typical error estimates due to these are summarized in Table 5
(see Papers I and II for the methods of estimation).
We adopt a conservative value of 10\% for \scal. 
The net error due to all noise sources, \sn, is given in columns (8), (9) and (10) of Table 5
for \nd, \td and F respectively.

We have examined the effect of these noise sources 
by carrying out a correlation analysis of the modified time series given by 

\begin{equation}
x_{mod} (i) ~ = ~ x_{obs} (i) ~ \pm ~ f \ \sn (i)
\hspace{1.0cm}
0 \le f \le 1
\end{equation}

\noindent
in which \xmod is taken 
to be a randomly selected value ranging from \xobs$-$\sn to \xobs+\sn with uniform probability.
The rank correlation coefficient is computed for all the three combinations (\bwtimemod, 
\bwfluxmod and \timefluxmod, where \ndm, \tdm and \Fm denote the new time series of \nd, \td and F
respectively) of this modified time series.
A large number ($ \sim $ 1000) of such time series have been analyzed, and the average value of all the
correlation coefficients is treated as the modified correlation coefficient, \rsm.
The results obtained in this manner are tabulated in columns (11), (12) and (13) of Table 5.
Comparison with the corresponding numbers in Table 3 shows that the reduction in degrees of correlations 
due to above-mentioned noise sources varies from a few percent to as large as 50\%.
This indicates that the underlying noise sources can be responsible for the reduced degrees of 
correlations (at least for some cases), but they cannot explain the lack of significant correlations 
($ ie., ~ | \rs | \la 0.1 $) or a reversal of the sense of correlation.
This is consistent with the fact that the modulations due to these noise sources are much smaller than 
the measured depths of modulations of the quantities.
Our analysis also shows the magnitudes of noise required for a substantial reduction in the strength of 
correlation, or a complete elimination of the existing correlation, are such that \sn $ \sim $ 50\%; 
this is illustrated in Fig. 4 using the example of PSR B0628$-$28 data.

We note that the more basic question of the noise required to reduce the theoretically expected correlations 
to the observed values cannot be precisely answered by this technique. 
Nevertheless, our method gives ample indication that poor correlations seen in the data cannot be 
accounted for by afore-mentioned noise sources alone.
If the degrees of correlations have been reduced to their present values due to some yet unrecognized 
noise sources, then they need to be such that the fractional uncertainties caused by them are much larger
than their modulation indices.
The presence of such noise sources seems to be quite unlikely.


\subsection{Flux Variations}

Pulsar flux variations can be broadly grouped under three categories, $viz$ (i) intrinsic variations, 
(ii) those due to DISS, and (iii) those due to RISS.
Intrinsic flux variations are mostly expected to take place on short time scales, with most pulsars showing
random fluctuations from pulse to pulse.
Almost all our data are taken over durations long enough to essentially quench such intrinsic fluctuations.
Most of our data were obtained during the hours when the ionosphere was not active; therefore, a
probable bias in the estimate of flux density due to ionospheric scintillations is not important.
Further, even if ionospheric scintillations were present with large modulations (typically 50\%) 
and over long time scales ($\sim$ 10 sec), there will be $\sim$ 700 independent cycles of fluctuations
during our typical observing scans ($ \sim $ 2 hours),
thereby reducing the bias to the level of $\sim$ 2\%.
The effect due to residual DISS fluctuations at any epoch is already included as a noise term,
as described in \S 5.1.
Poor flux correlations (\BF and \TF), as seen in our data, can result from two possibilities:
(i) RISS fluctuations of flux occurring on time scales different from those of \nd and \td, or
(ii) flux variations on time scales similar to refractive time scales, but caused by a hitherto
unidentified intrinsic source.

Regarding the first possibility, we note that though the theoretical models assume that \nd, \td and F vary
on similar time scales, there is no conclusive observational verification of the same.
Our data are not sampled regularly enough to obtain meaningful values for the fluctuation time scales of
these quantities using a correlation or structure function analysis.
In this context, it is interesting to note that most estimates of measured time scales for refractive flux
modulations (e.g. Gupta et al. 1993; Spangler et al. 1993) are reported to be substantially smaller than
the values predicted by theory.

Turning to the second possibility, we note that significant reductions in the flux correlations can result 
if, in addition to RISS modulations, there are large-amplitude intrinsic flux variations occurring over 
time scales comparable to refractive time scales.
Using the method described in \S 5.1, we have examined the effect on correlation properties of flux
density by treating the possible intrinsic flux variations in the form of a `noise source'.
The analysis shows that substantial amounts of intrinsic flux variations will be required to produce 
the poor flux correlations seen in the data.
The characteristics of such noise sources are found to be such that the fractional rms fluctuations 
due to them are quite comparable to the observed flux modulation indices (\mr).
Under such conditions, the RISS-induced flux variations will have to be unusually low
($ \sim $ 10$-$15\%).
A simple Kolmogorov form of density spectrum implied by such low values of flux modulation indices will be, 
however, in contradiction with the modulation characteristics of \nd and \td (denoted as 
\mb and \mt respectively), which suggest a spectrum steeper than $ \alpha = 11/3 $ (Paper II).
No theory predicts a situation of large-amplitude modulations of \nd and \td 
(\mb $ \sim $ 0.3$-$0.55 and \mt $ \sim $ 0.1$-$0.3) and small-amplitude flux modulations 
(\mr $ \sim $ 0.1$-$0.15).
Thus, it is unlikely that long-term intrinsic fluctuations (if they exist) are responsible for the 
reduced flux correlations seen in our data.


\subsection{Effect of Persistent Slopes of Patterns}

In our earlier papers (Papers I and II), we have discussed the anomalous scintillation behaviour of 
``persistent drift slopes'' seen with PSRs B0834+06 and B1919+21, where the drift slope of intensity
patterns in the dynamic spectra shows few or no sign reversals during the entire observing session.
We also described a classification scheme, based on the statistical characteristics of drift rate 
measurements, to distinguish between the cases of ``non-reversals'' and ``frequent reversals'' of 
drift slopes.
The data were accordingly categorized into Class II and Class I respectively, and those for which a clear 
distinction was not possible were categorized as `NC' (see column 6 Table 2).
The data from the first 3 observing sessions of PSR B0834+06 and the two sessions of 
PSR B1919+21 clearly come under Class II; there are 6 entries in `NC' and 14 in Class I.
Interestingly, we find none of the pulsars which shows reasonable agreement with the predictions 
belong to the special category, Class II.

Since the traditional decorrelation bandwidth (\nd) is affected by the presence of sloping patterns 
in dynamic spectra, the modulation characteristics of \nd can get significantly altered due to persistent
drift slopes.
Consequently, this may lead to significant changes in the correlation properties of \nd with the other 
quantities.  In Paper II, we defined a new quantity, ``drift-corrected decorrelation bandwidth'' (\ndc), 
which can be treated, to
first order, as \nd in the absence of refractive bending ($ie.,\ \refr = 0$).
Therefore, in the event of drifting bands playing a substantial role in the modulations of \nd,
it is reasonable to expect the correlations of \ndc to differ
significantly from that of \nd.

On carrying out a correlation analysis between the variations of \ndu and \td for PSR B0834+06,
for which the property of persistent drift slopes is extensively seen,
we find the results to be considerably different from that of the combination \nd and \td
(see Table 6).
Scatter plots illustrating this are shown in Figs. 5(a)$-$(h), where plots of \nd vs \td and \ndu vs \td 
are displayed for 4 sessions of PSR B0834+06 to highlight the change of trend on using \ndu.
Interestingly, the new correlations of \egtb and \egtc are in qualitative agreement with the prediction
for bandwidth-time correlation (Figs. 5.d and 5.f), although quantitative discrepancy prevails.
Also, for \egta, the method yields a considerable enhancement of the bandwidth-time correlation
$-$ \BT = 0.44 $ \longrightarrow $ \BcT = 0.59, which 
agrees with the theoretical value within the 90\% confidence limits (Figs. 5.a and 5.b).
No such improvement is seen for \egtd ({\it ie.,} \BT $ \approx $ \BcT) (Figs. 5.g and 5.h)
which shows drift reversals like many other pulsars.
Thus it appears that the method gives some meaningful results for data with persistent sloping patterns.

On applying this analysis to the data of another Class II pulsar, PSR B1919+21, 
we find that the improvements in the correlations are only marginal.
But we also note that these data are comparatively weaker examples of persistent drifts and are 
characterized by occasional sign reversals of drifts (Paper I).
For completeness, we carried out a similar analysis for data classified into other two categories,
the results from which are summarized in Figs. 6(a)$-$(c).
Here the correlation coefficients between the fluctuations of \ndc and \td (\BcT) are plotted
against those between the fluctuations of \nd and \td (\BT).
The changes in the correlation properties are found to be only marginal for Class I data, except for PSR
2327$-$20, which may be a spurious effect resulting from the systematic trend present in the time series of 
\td for this pulsar (Fig. 4.x of Paper I).
But among the data classified as `NC', PSR B2310+42 was found to be an exception where the 
method leads a significant enhancement in bandwidth-time correlation
(\BT = 0.04, whereas \BcT = 0.59), bringing it within reasonable agreement
with the predictions.
Interestingly, from the time series of drift rate measurements (Fig. 4.w of Paper I), 
it appears that this pulsar is similar to PSRs B0834+06 and B1919+21, 
though a clear distinction was not possible due to the poor statistical
quality of its data.

Figs. 6.b and 6.c summarize the other correlations of \ndc, $ie,$ \bwcflux and \bwcdrift.
In Fig. 6.b, \BcF is plotted against \BF.
Barring a few exceptions (indicated with the corresponding pulsar name alongside), we find the new
correlation coefficients to be only marginally different from those between the traditional decorrelation
bandwidth and flux density.
By and large, this seems to be in accordance with the simple expectations based on the existing RISS 
models, where the flux correlations are expected to arise from curvature effects, and, therefore, should 
not be affected by the drift slopes.
However, we note that, PSRs B0834+06 (I, II and III) and B2310+42, which show an improved correlation for
\bwctime, do not show a similar improvement for \bwcflux.
A plot of \BcD against \BD is shown in Fig. 6.c.
Values of the new correlation coefficients are consistently biased above the line of unity slope, which
means a significant reduction in the coefficient or a reversal of the sense. 
Such a behaviour is well in accordance with our presumption that the observed anti-correlation between
\nd and \dtn results from the reduction in \nd due to refractive gradients, and, hence, poorer 
correlations should result between the drift-corrected \nd and \dtn.

Although, the present analysis does not firmly establish the exact role of persistent drifts in 
modifying the correlation properties, it provides some evidence for refractive
modulations getting considerably modified in the presence of persistent drifts.
Modulation characteristics under such scenarios are not worked out by the available 
theoretical models for RISS.
In any case, persistent pattern drifts seem to be another plausible reason which can reduce the true 
correlations due to `normal RISS', thereby giving rise to inconsistencies with the predictions.


\subsection{Limitations and Possible Improvements of Theories}

The overall inconsistency of the present observations with theoretical predictions also raises 
the question of validity of some of the basic assumptions usually made by the theoretical models.
Here we address some of them which may be relevant for suitable refinements of the existing 
models.


\subsubsection{Effect of Extended and Inhomogeneous Media}

Most of the available theoretical calculations, including cross-correlation properties between scintillation
observables, have been confined to the simplest scenario of thin screen scattering models.
Romani et al. (1986) show that the thin screen theory underestimates the refractive
flux fluctuations by a factor $\sim$ 1.4 to 2.3 (depending on the power-law index of the density
irregularity spectrum) in comparison to a continuously distributed scattering medium.
Such detailed treatments are not available for the rest of the observables.
It is worth investigating how the correlation properties will be altered on considering more
realistic scattering geometries (such as multiple screens or a continuous medium).
Our data show that the measured flux density modulation indices are in better agreement with the theoretical
predictions of an extended scattering medium than those of a thin screen (Paper II).
Therefore, it is possible that the agreement between the observed correlation properties and the
predictions can also improve upon taking into consideration the effects due to extended medium.
Further, the extended geometry generally considered by the models is that of a homogeneous distribution 
of scattering material in the line of sight. 
Not much understanding exists on strong scattering effects due to a heterogeneous distribution of
scattering material.

The ISM is known to be of clumpy nature in general, and there are regions 
(such as supernova shocks, H {\sc ii} regions, etc) where the strength of scattering is much
larger than that in a typical region of the ISM.
Cordes et al. (1988, 1991) suggested the need for a clumped, intense component with a volume 
filling factor $ \approx ~ 10^{-4} ~ {\rm R_{pc}} $ (where $ {\rm R_{pc}} $ is the size of clump)
for the Galactic distribution of turbulence.
The strength of turbulence is believed to be $3-4$ orders of magnitude larger in this clump component.
However, since the present study has been restricted largely to close-by ($ \la $ 1 kpc) pulsars, 
scattering effects due to this component or those due to the Galactic spiral arms may not be important for
most of our pulsars.

Interestingly, our data have also shown evidence for a highly inhomogeneous 
distribution of scattering material in the LISM (Bhat et al. 1997, 1998), where a scattering structure 
of a bubble with a shell boundary is suggested for the solar neighborhood.
On examining the contributions to the scattering from different components, viz., the cavity
($ie.,$ interior of the bubble), the shell and the outer ISM, we find the lines-of-sight to the 
pulsars can be broadly categorized into four groups, $viz$; 
(i) predominant scattering ($ \ga $ 75\%) due to the outer ISM (Class A),
(ii) substantial contributions from both the shell as well as the outer ISM (Class B),
(iii) scattering due to the shell and the outer ISM, but predominantly ($ \ga $ 70\%) due to the former
(Class C) and
(iv) scattering due to the entire line-of-sight, in which the cavity also contributes 
significantly in addition to the shell and the outer ISM (Class D).
Classification of pulsars according to this scheme (column (7) of Table 2) shows that pulsars,
for which the observed correlations are in reasonable agreement with the predictions, 
do not come under any specific class.
However, we note that, in our data, the agreement with the predictions is generally seen for pulsars 
with comparatively larger DMs (20$-$35 \dmu).
While we are unable to identify any simple connection between the diversity of 
the correlation results and the structure of the LISM, 
it is plausible that the observed correlations are manifestations of some hitherto unrecognized 
ISS effects, presumably relevant in the case of heterogeneous media.

Over the past several years, observations have revealed various kinds of unusual scattering effects 
(such as multiple imaging events, ESEs and persistent drifts) which are attributed to the presence 
of large-scale dense refracting structures in the ISM
(e.g. Cordes \& Wolszczan 1986; Rickett, Lyne \& Gupta 1997; Fiedler et al. 1987, 1994; Gupta et al. 1994; 
Bhat et al. 1998b).
While details such as their possible associations with other Galactic structures and the distribution 
in the Galaxy still remain to be understood, the accumulated data give some indication of their important 
role in ISS of pulsars and radio sources. 
It is worth mentioning that two of these, ESEs and persistent drifts, occur over time scales comparable to
or longer than refractive time scales, and hence can potentially modify the time series of some of the 
scintillation observables, and may even alter the cross-correlations between their fluctuations. 
There is some evidence in support of this view from the present study and from 
Lestrade, Rickett \& Cognard (1998). 
While our analysis shows persistent drift slopes modifying the correlations between \nd, \td and \dtn (\S
5.3), Lestrade et al. (1998) discuss the effect of ESEs on the correlation between flux density and the 
pulse arrival time.
Recent paper by Clegg, Fey \& Lazio (1998) 
(also see Romani, Blandford \& Cordes 1987; Fiedler et al. 1994) analyze the
characteristics of flux variations due to discrete plasma lensing structures.
There are no theoretical treatments at present which address the perturbations on DISS observables 
due to such structures.
But there is some observational evidence to suggest that the role of discrete structures may be 
important in the cross-correlation properties of some of the observables.


\subsubsection{Refractive Effects and the Electron Density Spectrum}

As mentioned in \S 2, the cross-correlation properties of the scintillation observables have been worked out 
for simple power-law forms of density fluctuation spectra. Several attempts have been made in the recent past
to determine the exact form of the spectrum, but a conclusive picture is yet to emerge. There are conflicting
results from various kinds of measurements (cf. Paper II; Armstrong, Rickett \& Spangler 1995; Rickett 1990; 
Narayan 1988).
There are observations indicating that a simple power-law description is inadequate, and the spectrum needs 
to be more refractive in nature than a simple Kolmogorov form. Furthermore, there is substantial evidence 
(mainly from phenomena such as multiple imaging, ESEs and persistent drifts) suggesting the existence of 
localized dense refracting structures in the ISM, thereby favoring {\it non-power-law} forms of spectrum, 
at least for some lines of sight. Some viable options as indicated by our observations and several others 
from the literature are:
(i) a {\it composite } spectrum which steepens at low wavenumbers ($\sim 10^{-14}-10^{-11} \ {\rm m ^{-1}}$), 
(ii) a power-law form of spectrum (with $ \alpha \approx \kolind $) in combination with a separate
large-scale (say, $ \sim $ 10$-$100 AU) component, and
(iii) a power-law spectrum superposed with dense discrete structures.
According to Romani et al. (1986), if the density spectrum has a simple power-law form 
($ie,$ with unimportant cutoffs), then correlations between the fluctuations of \nd, \td and F
are qualitatively similar for different values of spectral slope ($\alpha$). But it is quite possible that, 
like several other refractive effects, correlation properties also turn highly sensitive to the spectral 
characteristics when the spectrum has a more complex form.
Theoretical developments in this direction remain to be made.


\subsubsection{Non-stationarity of the Medium}

Another observational result of interest is the evidence for an apparent lack of stability of the observed
correlation properties. 
We have four pulsars with multiple, well-separated observing sessions spanning $\sim$ 1 to 3 years, and
their correlation properties are found to be changing significantly from session to session.
The property is best illustrated by the data of PSR B0834+06, observations of which span a period of 
$\sim$ 1000 days (4 sessions), and substantial variations are seen between any two successive sessions.
In \S 4.3, we showed that this effect is not due to the statistical quality of the data
$-$ in terms of number of epochs of observation (\nep) and number of refractive cycles spanned (\nref).
Such an effect is unexpected if the ISM was well-behaved and the underlying density fluctuations were 
describable by a simple power-law spectrum.
We also note that the changes in correlation properties are sometimes accompanied by substantial changes in
the diffractive and/or refractive scintillation properties (Papers I and II).
While the observed variations of \nd and \td are explainable, in some cases, in terms of variations in the
scattering strength (\cn) or in the pattern velocity (\viss), it is not very obvious what kind of effects can 
lead to the variations seen in the correlation properties.
Long-term variations of this kind are difficult to understand in terms of simple models, and point to some 
hitherto unrecognized form of ISS or a more complex nature of the scattering medium.
If such variations are to be attributed to the medium, then the present observations do not support the 
assumption of a stationary medium.


\section{Conclusions}

We have analyzed data from our long-term pulsar observations to test the quantitative predictions given by
the theoretical models of refractive scintillation.
The data consist of dynamic scintillation spectra for 18 pulsars, which were regularly monitored at 10$-$90
epochs over time spans $\sim$ 100$-$1000 days.
They allow simultaneous measurements of the observables decorrelation bandwidth (\nd), 
scintillation time scale (\td), drift rate of the intensity patterns (\dtn) and 
flux density (F).
The observed fluctuations of these quantities are examined for their cross-correlation properties, and
compared with the existing predictions.
For 5 pulsars, there is reasonable agreement with the predictions, where a positive correlation between 
\nd and\td, and anti-correlations between \nd and F, and \td and F are seen.
The measured degrees of correlations are, however, generally lower than the predicted values.
While a number of data sets (roughly 60\%) show ``partial agreement'' 
($ie.,$ 1$-$2 combinations in qualitative agreement with the predictions, 
whilst an absence or an opposite correlation is the case with the remaining ones),
there are also examples (PSRs B0834+06(III) and B2045$-$16) for which the variations of all the 3
quantities are uncorrelated to each other.
A complexity of this kind is not easy to comprehend in terms of simple models.

Despite  inconsistency of the correlation results with the predictions, there are some general
trends.
A large number of pulsars show positive correlations between the fluctuations of decorrelation bandwidth
and scintillation time scale, while the disagreement is mainly due to the poor flux correlations.
Another interesting result is the anti-correlation seen between the fluctuations of decorrelation
bandwidth and drift slope for many pulsars.
Although the relevant quantitative predictions are not available at present, the observed behaviour 
is in accordance with the expectations based on the simple models of RISS.

The statistical quality of the data does not seem to be the cause of the inconsistency with
the theoretical predictions, as an improved
statistics (for pulsars with multiple observing sessions) does not result in a better agreement with the
predictions.
Our analysis shows, for part of the data, the underlying noise sources might be responsible for reduced
correlation coefficients.
Nevertheless, they cannot explain the absence of significant correlations seen
with a substantial (roughly one third) part of the data.
Our analysis suggests it is unlikely that the observed poor flux correlations arise from hitherto 
unrecognized large-amplitude intrinsic flux variations occurring on time scales similar to that of 
refractive fluctuations.
Further, we find the anomalous scintillation behaviour such as persistent drift slopes to be the cause 
of lack of correlations for part of the data (sessions II and III of PSR B0834+06, and PSR B2310+42), 
where the predicted positive correlation between \nd and \td turns up on using the drift-corrected \nd.

The qualitative agreement between the observed and predicted correlations for a number of pulsars 
indicates  that our basic  understanding of refractive scintillation is correct. 
However, reduced correlation coefficients and the absence of one or more 
of the predicted correlations suggests that in actual practice, conditions  
are more complex than assumed by the theoretical models. Some of the basic assumptions like
the  thin screen approximation and a simple power-law description of 
irregularity spectrum, need to be reexamined. The theory also needs to be enhanced to
take into account non-uniform distribution of scattering along the line-of-sight.
A more comprehensive theory incorporating these and additional features needs to be 
evolved to explain the observed complexities.
We hope the present observations will stimulate further theoretical work towards a better understanding of
various refractive scattering phenomena due to the ISM.


{\it Acknowledgments:}
The authors wish to thank J. Chengalur and R. T. Gangadhara for reading the manuscript and giving 
useful comments.
We thank our referee for several fruitful comments and suggestions towards improving the presentation of
our results in this paper.


{}

\clearpage


\begin{center} {\large\bf Figure Captions} \end{center}

\begin{description}

\item [Figure 1(a)$-$(l)]
Sample scatter plots for different combinations of the quantities \nd, \td and F, illustrating the
diversity seen in the correlation properties and their agreement with the predictions. 
The pulsar name (with session ID) is given at the top of each panel, and the rank correlation coefficient
(\rs) is indicated at the lower right corner.
The dotted lines indicate the mean values of the quantities.
The data of PSR B0823+26(II) is an example where all the 3 correlations are in reasonable agreement.
The agreement is only partial for PSR B0834+06(IV) and PSR B1919+21(II); two combinations are in agreement
for the former, whereas only a single combination agree with the prediction for the latter.
For PSR B2045$-$16, all the 3 quantities are uncorrelated to each other.

\item [Figure 2(a)$-$(c)]
Sample scatter plots illustrating the anti-correlation property between the variabilities of \nd and \dtn. 
The panel description is similar to Fig. 1.

\item [Figure 3(a)$-$(p)]
Correlation properties of 4 pulsars PSRs B0823+26, B0834+06, B1133+16 and B1919+21.
The data from different observing sessions are combined to improve the statistical 
quality and reliability of the data (in terms of \nep and \nref).
The panel description is similar to Fig. 1.

\item [Figure 4]
Effect of noise on the correlation properties, illustrated using the data of PSR B0628$-$28.
Modified correlation coefficients (\rsm) are plotted against the magnitude of noise introduced (\sn)
(in quantities \nd, \td and F) for the combinations \bwtimemod, \bwfluxmod and \timefluxmod 
Different symbols used are explained at the top right corner.

\item [Figure 5(a)$-$(h)]
Effect of persistent drifting bands on the bandwidth-time correlation of PSR B0834+06.
The panel description is similar to Fig. 1.
There are two panels each for the data from a given session, the first one is the scatter plot of
traditional \nd against \td, while the second one is for the drift-corrected \nd (\ndc) against \td.

\item [Figure 6(a)$-$(c)]
Results of the correlation analysis of drift-corrected decorrelation bandwidth \ndc.
Panel (a): correlation coefficients between \ndc and \td are plotted against those between \nd and \td,
panel (b): correlation coefficients between \ndc and F are plotted against those between \nd and F, and 
panel (c): correlation coefficients between \ndc and \dtn are plotted against those between \nd and \dtn. 
The dashed line is of unity slope, and dotted lines define a band of 10\% discrepancy.
Different symbols used are explained inside the figure itself. 

\end{description}



\begin{thebibliography}{}

\bibitem{} Armstrong, J. W., Rickett, B. J. \& Spangler, S. R. 1995, ApJ, 443, 209
\bibitem{} Bhat, N. D. R., Gupta, Y. \& Rao, A. P. 1997,
in Proceedings of the IAU Colloquium No. 166 ``The Local Bubble and Beyond'',
eds. D. Breitschwerdt, M. J. Freyberg, J. Tr\"umper,
Lecture Notes in Physics 506, 211
(Springer-Verlag)
\bibitem{} Bhat, N. D. R., Gupta, Y. \& Rao, A. P. 1998, ApJ, 500, 262
\bibitem{} Bhat, N. D. R., Rao, A. P. \& Gupta, Y. 1998a, In Press (Paper I)
\bibitem{} Bhat, N. D. R., Gupta, Y. \& Rao, A. P. 1998b, In Press (Paper II)
\bibitem{} Blandford, R. D. \& Narayan, R. 1985, MNRAS, 213, 591
\bibitem{} Bondi, M., Padrielli, L., Gregorini, L., Mantovani, F., Shapirovskaya, N. \& Spangler, S. R.
1994, A\&A, 287, 390
\bibitem{} Clegg, A. W., Fey, A. L. \& Lazio, T. J. 1998, ApJ, 496, 253
\bibitem{} Cordes, J. M., Pidwerbetsky, A. \& Lovelace, R. V. E. 1986, ApJ, 310, 737
\bibitem{} Cordes, J. M., Rickett, B. J. \& Backer, D. C. 1988, 
AIP Conf. Proc. No. 174 - Radiowave Scattering in the Interstellar Medium
(New York:AIP)
\bibitem{} Cordes, J. M., Spangler, S. R., Weisberg, J. M. \& Clifton, T. R. 1988,
in AIP Conf. Proc. No. 174 - Radiowave Scattering in the Interstellar Medium,
ed. Cordes, J. M., Rickett, B. J. \& Backer, D. C.
(New York: AIP), 180
\bibitem{} Cordes, J. M., Weisberg, J. M., Frail, D. A., Spangler, S. R. \& Ryan, M. 1991, Nature, 354, 121
\bibitem{} Cordes, J. M. \& Wolszczan 1986, ApJ, 307, L27
\bibitem{} Diaconis, P. \& Efron, B. 1983, Sci. Am., 248, no. 5, 116
\bibitem{} Efron, B. 1979, Anns. Stat., 7, 1
\bibitem{} Fiedler, R. L., Dennison, B., Johnston, K. J. \& Hewish, A. 1987, Nature, 326, 675
\bibitem{} Fiedler, R. L., Dennison, B., Johnston, K. J., Waltman, E. B. \& Simon, R. S. 1994, ApJ, 430,
581
\bibitem{} Hewish, A. 1992, Phil. Trans. Royal Soc., 341, 167
\bibitem{} Gould, D. M. 1994, Ph.D. thesis, University of Manchester
\bibitem{} Gupta, Y., Rickett, B. J. \& Coles, W. A. 1993, ApJ, 403, 183
\bibitem{} Gupta, Y., Rickett, B. J. \& Lyne, A. G. 1994, MNRAS, 269, 1035
\bibitem{} Kaspi, V. M. \& Stinebring, D. R. 1992, ApJ, 392, 530
\bibitem{} Lestrade, J., Cognard, I. \& Biraud, F. 1995, 
in Millisecond Pulsars: A Decade of Surprise, 
eds. Fruchter, A., Tavani, M. \& Backer, D. 
(San Francisco: Astronomical Society of the Pacific Conference Series), 375
\bibitem{} Lestrade, J., Rickett, B. J. \& Cognard, I. 1998, A\&A, 334, 1068
\bibitem{} Narayan, R. 1992, Phil. Trans. Royal Soc., 341, 151
\bibitem{} Press, W. H., Flannery, B. P., Teukolsky, S. A. \& Vetterling, W. T. 1992
Numerical Recipes - The Art of Scientific Computing (Cambridge: Cambridge University Press)
\bibitem{} Rickett, B. J. 1990, ARA\&A, 28, 561
\bibitem{} Rickett, B. J., Coles, W. A. \& Bourgois G. 1984, A\&A, 134, 390
\bibitem{} Rickett, B. J., Lyne, A. G. \& Gupta, Y. 1997, MNRAS, 287, 739
\bibitem{} Romani, R. W., Blandford, R. D. \& Cordes, J. M. 1987, Nature, 328, 324
\bibitem{} Romani, R. W., Narayan, R. \& Blandford, R. D. 1986, MNRAS, 220, 19
\bibitem{} Sieber, W. 1982, A\&A, 113, 311
\bibitem{} Spangler, S. R., Eastman, W. A., Gregorini, L., et al. 1993, A\&A, 267, 213
\bibitem{} Stinebring, D. R. \& Condon, J. J. 1990, ApJ, 352, 207
\bibitem{} Stinebring, D. R., Faison, M. D. \& McKinnon, M. M. 1996, ApJ, 460, 460
\bibitem{} Wolszczan, A. \& Cordes, J. M. 1987, ApJ, 320, L35

\end{thebibliography}
\end{document}


\newcommand{\bwtime}{$ \{ \nu _d , \tau _d \} $\,}
\newcommand{\bwflux}{$ \{ \nu _d , F \} $\,}
\newcommand{\timeflux}{$ \{ \tau _d , F \} $\,}

\begin{deluxetable}{cccc}
\tablenum{1}
\tablewidth{0pt}
\tablecaption{Predicted Correlation Coefficients from Romani et al. (1986)}
\tablehead{
Powrer-law Index & Decorrelation bandwidth & Decorrelation bandwidth & Scintillation time scale \\
                 & and			   & and		     & and                      \\
                 & Scintillaton time scale & Flux Density            & Flux Density             \\
$ \alpha $       & \bwtime                 & \bwflux                 & \timeflux                \\ \\
(1)              & (2)                     & (3)                     & (4)                      \\
}
\startdata
                   &      &           &           \\
$ { 11 \over 3 } $ & 0.75 & $ -0.76 $ & $ -0.50 $ \\
                   &      &           &           \\
4                  & 0.77 & $ -0.80 $ & $ -0.58 $ \\
                   &      &           &           \\
4.3                & 0.79 & $ -0.84 $ & $ -0.64 $ \\
                   &      &           &           \\
\enddata
\end{deluxetable}

\newcommand{\nep}{\mbox{$ {\rm N_{ep} } $ }}
\newcommand{\tsp}{\mbox {$ {\rm T_{sp} } $ }}

\begin{deluxetable}{cllcccc}
\tablenum{2}
\tablewidth{0pt}
\tablecaption{Pulsar Sample: Observing Parameters and Classification}
\tablehead
{
No. & Pulsar (Session) & Period of & \nep & \tsp & Drift Class$^{\dagger}$ & LOS Class$^{\ddagger}$ \\
    &                  & Observation &    & (days) &                       &                        \\ \\
(1) & (2)	       & (3)	   & (4)  & (5)  & (6)			   & (7)		    \\
}
\startdata
01 & PSR B$0329+54$  	 & Apr 95$-$Jul 95 & 14 &  66 & NC & A \\
02 & PSR B$0628-28$  	 & Oct 93$-$Jan 94 & 17 &  83 & I  & B \\
03 & PSR B$0823+26$(I)	 & Mar 93$-$May 93 & 12 &  66 & I  & C \\
04 & PSR B$0823+26$(II)	 & Oct 93$-$Jan 94 & 19 &  87 & I  &---\\
05 & PSR B$0834+06$(I)	 & Jan 93$-$May 93 & 27 & 120 & II & C \\
06 & PSR B$0834+06$(II)	 & Oct 93$-$Jan 94 & 18 &  74 & II &---\\
07 & PSR B$0834+06$(III) & Feb 94$-$Jun 94 & 21 &  78 & II &---\\
08 & PSR B$0834+06$(IV)	 & Apr 95$-$Jul 95 & 27 & 100 & I  &---\\
09 & PSR B$0919+06$  	 & Mar 94$-$Jun 94 & 19 &  70 & I  & C \\
10 & PSR B$1133+16$(I)	 & Jan 93$-$Feb 93 & 06 &  19 & I  & D \\ 
11 & PSR B$1133+16$(II)	 & Feb 94$-$Jun 94 & 25 &  90 & I  &---\\ 
12 & PSR B$1133+16$(III) & Apr 95$-$Jul 95 & 27 &  95 & I  &---\\
13 & PSR B$1237+25$  	 & Oct 93$-$Jan 94 & 09 &  77 & I  & D \\
14 & PSR B$1508+55$  	 & Apr 95$-$Jul 95 & 09 &  52 & I  & B \\
15 & PSR B$1540-06$  	 & Apr 95$-$Jul 95 & 12 &  91 & NC & B \\
16 & PSR B$1604-00$  	 & Apr 95$-$Jul 95 & 10 &  93 & I  & B \\
17 & PSR B$1747-46$  	 & Apr 95$-$Jul 95 & 12 &  92 & I  & A \\
18 & PSR B$1919+21$(I)	 & Mar 93$-$May 93 & 15 &  59 & II & B \\
19 & PSR B$1919+21$(II)	 & Oct 93$-$Jan 94 & 48 &  87 & II &---\\
20 & PSR B$1929+10$  	 & Mar 94$-$Jun 94 & 09 &  42 & I  & C \\
21 & PSR B$2016+28$  	 & Oct 93$-$Jan 94 & 20 &  83 & NC & A \\
22 & PSR B$2020+28$  	 & Mar 94$-$Jun 94 & 15 &  74 & I  & A \\
23 & PSR B$2045-16$  	 & Oct 93$-$Jan 94 & 35 &  85 & I  & B \\
24 & PSR B$2310+42$  	 & Apr 95$-$Jul 95 & 10 &  77 & NC & A \\
25 & PSR B$2327-20$  	 & Mar 93$-$Jun 94 & 18 &  68 & I  & C \\
\tablenotetext{\dagger}{Based on the statistical characteristics of drift rate measurements (see Paper II for details).}
\tablenotetext{\ddagger}{Based on the scattering geometry of the line-of-sight (LOS) expected from the model
for the local interstellar medium (Bhat et al. 1998).}
\enddata
\end{deluxetable}

\renewcommand{\baselinestretch}{1.65}

\newcommand{\rs}{\mbox{$ r_s $}}
\newcommand{\nd}{\mbox{$ \nu _d $}}
\newcommand{\td}{\mbox{$ \tau _d $}}
\newcommand{\dtn}{\mbox{$ d t / d \nu $}}
\newcommand{\BT}{$ r_s \{ \nu _d , \tau _d \} $\,}
\newcommand{\BF}{$ r_s \{ \nu _d , F \} $\,}
\newcommand{\TF}{$ r_s \{ \tau _d , F \} $\,}
\newcommand{\BD}{$ r_s \{ \nu _d , d t / d \nu \} $\,}

\addtolength{\topmargin}{-4.5cm}
\small

\begin{deluxetable}{clrrrrc}
\tablenum{3}
\tablewidth{0pt}
\tablecaption{Results from the Correlation Analysis}
\tablehead
{
Sl. No. & Pulsar (Session) & \BT              & \BF            & \TF            & \BD               & Remarks$^{\dagger}$ \\
(1)     & (2)              & (3)              & (4)            & (5)            & (6)               & (7) 
}
\startdata
01 & PSR B$0329+54$      &${\rm 0.04_{-0.59}^{+0.45}}$ &${\rm 0.45_{-0.31}^{+0.72}}$ &${\bf-0.61_{-0.94}^{-0.26}}$ &${\rm-0.01_{-0.70}^{+0.50}}$ & 1A \\
02 & PSR B$0628-28$      &${\bf 0.52_{-0.04}^{+0.75}}$ &${\bf-0.36_{-0.82}^{+0.00}}$ &${\bf-0.70_{-0.95}^{-0.44}}$ &${\sf-0.41_{-0.90}^{+0.05}}$ & RA \\
03 & PSR B$0823+26$(I)   &${\bf 0.47_{-0.25}^{+0.79}}$ &${\rm 0.09_{-0.59}^{+0.41}}$ &${\rm-0.03_{-0.62}^{+0.29}}$ &${\sf-0.38_{-0.90}^{+0.05}}$ & 1A \\
04 & PSR B$0823+26$(II)  &${\bf 0.60_{-0.04}^{+0.76}}$ &${\bf-0.29_{-0.71}^{+0.09}}$ &${\bf-0.29_{-0.79}^{+0.05}}$ &${\sf-0.55_{-0.89}^{-0.26}}$ & RA \\
05 & PSR B$0834+06$(I)   &${\bf 0.44_{+0.09}^{+0.66}}$ &${\rm 0.28_{+0.00}^{+0.50}}$ &${\rm 0.37_{+0.07}^{+0.59}}$ &${\sf-0.18_{-0.50}^{+0.12}}$ & 1A \\
06 & PSR B$0834+06$(II)  &${\rm-0.14_{-0.58}^{+0.22}}$ &${\bf-0.24_{-0.62}^{+0.12}}$ &${\rm-0.04_{-0.42}^{+0.29}}$ &${\sf-0.69_{-0.90}^{-0.48}}$ & 1A \\
07 & PSR B$0834+06$(III) &${\rm 0.07_{-0.44}^{+0.44}}$ &${\rm-0.07_{-0.44}^{+0.23}}$ &${\rm-0.02_{-0.44}^{+0.32}}$ &${\sf-0.30_{-0.68}^{+0.09}}$ & 0A \\
08 & PSR B$0834+06$(IV)  &${\bf 0.42_{+0.01}^{+0.65}}$ &${\bf-0.23_{-0.53}^{+0.05}}$ &${\rm 0.30_{+0.00}^{+0.50}}$ &${\sf-0.39_{-0.71}^{-0.06}}$ & 2A \\
09 & PSR B$0919+06$      &${\bf 0.60_{+0.25}^{+0.75}}$ &${\bf-0.21_{-0.68}^{+0.17}}$ &${\bf-0.19_{-0.65}^{+0.17}}$ &${\sf-0.20_{-0.59}^{+0.17}}$ & RA \\
10 & PSR B$1133+16$(I)   &${\bf 0.42_{+0.19}^{+0.88}}$ &${\bf-0.37_{-1.00}^{+0.17}}$ &${\rm-0.09_{-0.97}^{+0.34}}$ &${\sf-0.43_{-1.00}^{+0.22}}$ & 2A \\
11 & PSR B$1133+16$(II)  &${\bf 0.83_{+0.62}^{+0.91}}$ &${\rm-0.04_{-0.41}^{+0.25}}$ &${\rm 0.01_{-0.34}^{+0.32}}$ &${\sf-0.37_{-0.68}^{-0.08}}$ & 1A \\
12 & PSR B$1133+16$(III) &${\bf 0.32_{-0.07}^{+0.57}}$ &${\rm-0.01_{-0.38}^{+0.29}}$ &${\rm 0.09_{-0.25}^{+0.35}}$ &${\sf-0.67_{-0.81}^{-0.50}}$ & 1A \\
13 & PSR B$1237+25^*$    &${\bf 0.43_{-0.35}^{+0.66}}$ &${\rm 0.86_{+0.32}^{+0.97}}$ &${\rm 0.07_{-0.70}^{+0.44}}$ &${\sf-0.50_{-0.96}^{-0.04}}$ & 1A \\
14 & PSR B$1508+55$      &${\bf 0.40_{-0.48}^{+0.62}}$ &${\bf-0.57_{-0.97}^{-0.02}}$ &${\bf-0.86_{-1.00}^{-0.75}}$ &${\sf-0.38_{-0.97}^{+0.28}}$ & RA \\
15 & PSR B$1540-06^*$    &${\rm 0.05_{-0.62}^{+0.47}}$ &${\rm 0.43_{-0.20}^{+0.65}}$ &${\rm-0.07_{-0.79}^{+0.41}}$ &${\sf-0.56_{-0.94}^{-0.23}}$ & SC \\
16 & PSR B$1604-00^*$    &${\bf 0.22_{-0.52}^{+0.53}}$ &${\rm-0.02_{-0.63}^{+0.34}}$ &${\bf-0.31_{-0.88}^{+0.16}}$ &${\sf-0.71_{-0.98}^{-0.41}}$ & 2A \\
17 & PSR B$1747-46$      &${\bf 0.39_{-0.22}^{+0.68}}$ &${\rm+0.31_{-0.55}^{+0.75}}$ &${\rm 0.32_{-0.45}^{+0.55}}$ &${\rm 0.15_{-0.41}^{+0.41}}$ & 1A \\
18 & PSR B$1919+21$(I)   &${\bf 0.21_{-0.28}^{+0.47}}$ &${\rm 0.01_{-0.48}^{+0.35}}$ &${\rm-0.05_{-0.56}^{+0.34}}$ &${\sf-0.32_{-0.75}^{+0.08}}$ & 1A \\
19 & PSR B$1919+21$(II)  &${\bf 0.39_{+0.15}^{+0.58}}$ &${\rm 0.18_{-0.08}^{+0.41}}$ &${\rm 0.14_{-0.10}^{+0.35}}$ &${\sf-0.57_{-0.71}^{-0.36}}$ & 1A \\
20 & PSR B$1929+10^*$    &${\rm 0.10_{-0.81}^{+0.41}}$ &${\rm 0.55_{-0.48}^{+0.78}}$ &${\rm 0.19_{-0.66}^{+0.42}}$ &${\sf-0.33_{-0.96}^{+0.12}}$ & SC \\
21 & PSR B$2016+28^*$    &${\bf 0.53_{+0.17}^{+0.71}}$ &${\bf-0.13_{-0.58}^{+0.31}}$ &${\rm 0.30_{-0.22}^{+0.69}}$ &${\sf-0.69_{-0.90}^{-0.44}}$ & 2A \\
22 & PSR B$2020+28$      &${\bf 0.45_{+0.01}^{+0.64}}$ &${\bf-0.56_{-0.85}^{-0.02}}$ &${\bf-0.66_{-0.89}^{-0.17}}$ &${\rm-0.03_{-0.58}^{+0.47}}$ & RA \\
23 & PSR B$2045-16$      &${\rm 0.03_{-0.25}^{+0.29}}$ &${\rm 0.08_{-0.22}^{+0.32}}$ &${\rm 0.00_{-0.35}^{+0.31}}$ &${\sf-0.54_{-0.75}^{-0.31}}$ & 0A \\
24 & PSR B$2310+42^*$    &${\rm 0.04_{-0.68}^{+0.56}}$ &${\bf-0.67_{-0.98}^{-0.28}}$ &${\bf-0.15_{-0.88}^{+0.32}}$ &${\rm+0.13_{-0.76}^{+0.77}}$ & 1A \\
25 & PSR B$2327-20$      &${\rm-0.04_{-0.68}^{+0.38}}$ &${\rm 0.27_{-0.35}^{+0.61}}$ &${\rm 0.40_{+0.00}^{+0.70}}$ &${\rm+0.06_{-0.56}^{+0.52}}$ & SC \\
\tablenotetext{\dagger}{On the agreement with the predictions given in Table 1.}
\tablenotetext{*}{Correlations may not be reliable, as described in \S 3.0.}
\tablenotetext{ }{Notes: RA $-$ Reasonable Agrrement, 2A, 1A and 0A $-$ 2, 1 and 0 
combination(s) is(are) in Agreement,}
\tablenotetext{ }{SC $-$ Spurious Correlations.}
\enddata
\end{deluxetable}

\renewcommand{\baselinestretch}{1.25}

\addtolength{\topmargin}{+3.0cm}

\begin{deluxetable}{clrrrr}
\tablenum{4}
\tablewidth{0pt}
\tablecaption{Correlation results of the combined data$^{\dagger}$}
\tablehead
{
No. & \multicolumn{1}{c}{Pulsar} & 
\multicolumn{1}{c}{\BT} & \multicolumn{1}{c}{\BF} & \multicolumn{1}{c}{\TF} & \multicolumn{1}{c}{\BD} \\ \\
(1) & \multicolumn{1}{c}{(2)} & \multicolumn{1}{c}{(3)} & \multicolumn{1}{c}{(4)} & \multicolumn{1}{c}{(5)} &
\multicolumn{1}{c}{(6)} \\
}
\startdata
   &	&	&	&	&	\\
01 & PSR B0823+26 &${\bf +0.57 _{+0.25} ^{+0.75}}$ &${\rm +0.12 _{-0.19} ^{+0.38}}$&${\rm +0.28 _{-0.10} ^{+0.55}}$& 
${\sf -0.33 _{-0.62} ^{-0.05}}$ \\
   &	&	&	&	&	\\
02 & PSR B0834+06 &${\bf +0.27 _{+0.08} ^{+0.44}}$ &${\rm +0.00 _{-0.17} ^{+0.16}}$&${\rm +0.33 _{+0.17} ^{+0.47}}$&
${\sf -0.49 _{-0.64} ^{-0.34}}$ \\
   &	&	&	&	&	\\
03 & PSR B1133+16 &${\bf +0.58 _{+0.14} ^{+0.77}}$ &${\rm +0.04 _{-0.23} ^{+0.26}}$&${\rm +0.07 _{-0.14} ^{+0.28}}$&
${\sf -0.49 _{-0.65} ^{-0.25}}$ \\
   &	&	&	&	&	\\
04 & PSR B1919+21 &${\rm -0.10 _{-0.31} ^{+0.12}}$ &${\rm -0.12 _{-0.35} ^{+0.11}}$&${\rm +0.23 _{+0.00} ^{+0.41}}$& 
${\sf -0.69 _{-0.79} ^{-0.56}}$ \\
   &	&	&	&	&	\\
\tablenotetext{\dagger}{For pulsars with multiple observing sessions.}
\enddata
\end{deluxetable}

\setlength{\oddsidemargin}{-1.0cm}

\newcommand{\delvep}{\mbox{ $ \Delta V_{obs _{\bot }} $}}
\newcommand{\bwtimemod}{$ \{ \nu _d ^m , \tau _d ^m \} $\,}
\newcommand{\bwfluxmod}{$ \{ \nu _d ^m , F ^m \} $\,}
\newcommand{\timefluxmod}{$ \{ \tau _d ^m , F ^m \} $\,}

\addtolength{\oddsidemargin}{-0.75cm}
\addtolength{\topmargin}{-1.00cm}

\begin{deluxetable}{clrrrrrrrrrrr}
\tablenum{5}
\tablewidth{0pt}
\tablecaption{Effects of noise sources on correlation results}
\tablehead
{
No. & \multicolumn{1}{c}{Pulsar} & 
$\sigma _{mod}^{\dagger}$ & $\sigma _{est}^{\ddagger}$ & $\sigma _{pol}^@$ & $\sigma _{v,obs}^{\star}$ & 
$\sigma _{v,irr}^{\ast}$ & \multicolumn{3}{c}{$ \sigma _n $} & \multicolumn{3}{c}{$r_s^m$} \\ \\
 & & & & & & & ($ \nu _d $)  &  ($ \tau _d $) &  (F) & \bwtimemod & \bwfluxmod & \timefluxmod \\ \\
 & & (\%) & (\%) & (\%) & (\%) & (\%) & (\%) & (\%) & (\%) & & & \\ \\
(1) & \multicolumn{1}{c}{(2)} & (3) & (4) & (5) & (6) & (7) & (8) & (9) & (10) & (11) & (12) & (13) \\
}
\startdata
01 & PSR B$0329+54$   	& ~8 & ~5 &~0&~3&~5&~9&11&11& $ ....$&$ ....$&$-0.46$ \\
02 & PSR B$0628-28$   	& ~9 & ~6 &~2&~3&~6&11&13&12& $ 0.26$&$-0.29$&$-0.41$ \\
03 & PSR B$0823+26$(I) 	& 14 & ~5 &18&~5&~4&15&16&21& $ 0.42$&$ ....$&$ ....$ \\
04 & PSR B$0823+26$(II) & 14 & ~4 &18&10&~4&15&18&21& $ 0.22$&$-0.14$&$-0.21$ \\
05 & PSR B$0834+06$(I) 	& ~6 & ~9 &~1&14&~6&11&19&13& $ 0.31$&$ 0.29$&$ 0.28$ \\
06 & PSR B$0834+06$(II) & ~6 & ~7 &~1&12&~5&~9&16&12& $-0.10$&$-0.18$&$ ....$ \\
07 & PSR B$0834+06$(III)& ~6 & ~6 &~1&~5&~4&~8&11&12& $ ....$&$ ....$&$ ....$ \\
08 & PSR B$0834+06$(IV) & ~6 & ~7 &~1&~9&~4&~9&13&12& $ 0.31$&$-0.20$&$ 0.14$ \\
09 & PSR B$0919+06$   	& 16 & ~5 &14&~2&~2&17&17&18& $ 0.44$&$-0.11$&$-0.18$ \\
10 & PSR B$1133+16$(I)  & ~7 & ~7 &10&~6&~3&10&12&16& $ 0.80$&$-0.25$&$ ....$ \\
11 & PSR B$1133+16$(II) & ~7 & ~7 &10&~6&~3&10&12&16& $ 0.71$&$ ....$&$ ....$ \\
12 & PSR B$1133+16$(III)& ~7 & ~6 &10&~4&~3&~9&10&15& $ 0.20$&$ ....$&$ ....$ \\
13 & PSR B$1237+25$   	& ~7 & 11 &38&~4&~3&13&14&41& $ 0.35$&$ 0.82$&$ 0.27$ \\
14 & PSR B$1508+55$   	& 17 & ~4 &~1&~0&~2&17&18&11& $ 0.28$&$-0.44$&$-0.65$ \\
15 & PSR B$1540-06$   	& 11 & ~5 &17&24&13&12&30&20& $ ....$&$ 0.44$&$ ....$ \\
16 & PSR B$1604-00$   	& ~5 & 10 &~5&24&16&11&31&15& $ 0.09$&$ ....$&$-0.16$ \\
17 & PSR B$1747-46$   	& 13 & ~4 &10&~3&~5&14&15&15& $ 0.21$&$ ....$&$ ....$ \\
18 & PSR B$1919+21$(I) 	& ~5 & ~8 &18&~2&~8&~9&13&22& $ 0.13$&$ ....$&$ ....$ \\
19 & PSR B$1919+21$(II) & ~5 & ~9 &18&~5&~5&10&12&22& $ 0.13$&$ 0.18$&$-0.06$ \\
20 & PSR B$1929+10$   	& ~6 & 16 &49&~6&~7&17&19&53& $ ....$&$ 0.53$&$-0.08$ \\
21 & PSR B$2016+28$   	& ~6 & 12 &~2&14&18&13&26&16& $ 0.24$&$-0.12$&$ 0.11$ \\
22 & PSR B$2020+28$   	& ~8 & ~5 &~1&~2&~5&~9&11&11& $ 0.23$&$-0.22$&$-0.28$ \\
23 & PSR B$2045-16$   	& 12 & ~6 &~4&~5&~2&13&14&12& $ ....$&$ ....$&$ ....$ \\
24 & PSR B$2310+42$   	& 13 & ~5 &~4&~4&11&14&18&12& $ ....$&$-0.62$&$ ....$ \\
25 & PSR B$2327-20$   	& ~4 & ~8 &~3&22&13&~9&27&13& $ ....$&$ 0.07$&$ 0.27$ \\
\tablenotetext{\dagger} {Typical errors on the quantities $ \nu _d $ and $ \tau _d $.}
\tablenotetext{\ddagger} {Typical errors on the quantities $ \nu _d $, $ \tau _d $ and F.}
\tablenotetext{@} {The ionospheric contribution to the rotation measure is assumed to 
be $\sim$ 1 $ {\rm rad ~ m ^{-2}} $.}
\tablenotetext{\star} {For the net change (\delvep) in the transverse component of 
Earth's orbital motion during the time span of observation.}
\tablenotetext{\ast} {On assuming a motion of $\sim$ 10 $ {\rm km ~ secs ^{-1}} $ 
(Bondi et al. 1994) for the density irregularities.}
\enddata
\end{deluxetable}

\addtolength{\oddsidemargin}{+0.75cm}

\newcommand{\corrbwtime}{\mbox{$ r_s \{ \nu _{d_c} , \tau _d \} $\,}}

\begin{deluxetable}{lcc}
\tablenum{6}
\tablewidth{0pt}
\tablecaption{Correlation properties of the drift-corrected decorrelation bandwidth}
\tablehead
{
     & Decorrelation bandwidth  & Corrected Decorrelation bandwidth \\
     & and 		        & and 				   \\
Data & Scintillation time scale & Scintillation time scale          \\ \\
     & \bwtime & \corrbwtime \\ 
}
\startdata
                 	&                              &                            \\
PSR B$0834+06$(I) 	&${\bf  +0.44  _{+0.09} ^{+0.66}} $  & ${\bf +0.59 _{+0.28} ^{+0.75}} $ \\
                 	&                              &                            \\
PSR B$0834+06$(II) 	&${\rm  -0.14  _{-0.58} ^{+0.22}} $  & ${\bf +0.35 _{-0.08} ^{+0.62}} $ \\ 
                 	&                              &                            \\
PSR B$0834+06$(III) 	&${\rm  +0.07  _{-0.44} ^{+0.44}} $  & ${\bf +0.29 _{-0.14} ^{+0.55}} $ \\
                 	&                              &                            \\
PSR B$0834+06$(IV) 	&${\bf  +0.42  _{+0.01} ^{+0.65}} $  & ${\bf +0.43 _{+0.06} ^{+0.65}} $ \\
                 	&                              &                            \\
PSR B$2310+42$ 		&${\rm  -0.04  _{-0.68} ^{+0.56}} $  & ${\bf +0.59 _{-0.25} ^{+0.79}} $ \\
                 	&                              &                            \\
\enddata
\end{deluxetable}